\documentclass{aa}
\usepackage{natbib}
\bibpunct{(}{)}{;}{a}{}{,}
\begin{document}

   \title{Molecules as magnetic probes of starspots}

   \author{N. Afram\inst{1}
          \and S.V.~Berdyugina\inst{1,2}
          }

   \institute{Kiepenheuer Institut f\"ur Sonnenphysik, Freiburg, Germany 
    \and NASA Astrobiology Institute, Institute for Astronomy, University of Hawaii, USA}
           
   \date{Received date; accepted date}

  \abstract
   {Stellar dynamo processes can be explored by measuring the
magnetic field. This is usually obtained using the atomic and molecular Zeeman effect in spectral lines. While the atomic Zeeman effect can only access warmer regions, the use of molecular lines is of advantage for studying cool objects. The molecules MgH, TiO, CaH, and FeH are suited to probe stellar magnetic fields, each one for a different range of spectral types, by considering the signal that is obtained from modeling various spectral types.}
   {We have analyzed the usefulness of different molecules (MgH, TiO, CaH, and FeH) as diagnostic tools for studying stellar magnetism on active G-K-M dwarfs. We investigate the temperature range in which the selected molecules can serve as indicators for magnetic fields on highly active cool stars and present synthetic Stokes profiles for the modeled spectral type.}
   {We modeled a star with a spot size of 10\% of the stellar disk and a spot comprising either only longitudinal or only transverse magnetic fields and estimated the strengths of the polarization Stokes $V$ and $Q$ signals for the molecules MgH, TiO, CaH, and FeH. We combined various photosphere and spot models according to realistic scenarios.}
   {In G dwarfs, the  molecules MgH and FeH show overall the strongest Stokes $V$ and $Q$ signals from the starspot, whereas FeH has a stronger Stokes $V$ signal in all G dwarfs, with a spot temperature of 3800K.  In K dwarfs, CaH signals are generally stronger, and the TiO signature is most prominent in M dwarfs. }
   {Modeling synthetic polarization signals from starspots for a range of G-K-M dwarfs leads to differences in the prominence of various molecular signatures in different wavelength regions, which helps to efficiently select targets and exposure times for observations.}


   \maketitle

\section{Introduction}

\begin{table*}[!ht]
\caption{General comparison of the obtained signals showing the advantages of different molecules for various spectral type investigations. Calculations were made with various combinations of underlying models for each spectral type. We list the minimum residual intensity (I), Stokes $V$, and Stokes $Q$ signals from the different molecular lines for a model star with vsini=10~km/s and 10\% of the longitudinal or transverse magnetic field of 3 kG. Tabulated maxima exclude excursions due to atomic lines that overlap the relevant molecular band.}
\label{tab:overview}
\centering
\begin{tabular}{c|c|rrrr|rrrr|rrrr}
\hline
spectral type &$T_{\rm phot}$/$T_{\rm spot}$ & \multicolumn{4}{c|}{Minimum residual intensity}& \multicolumn{4}{c|}{Maximum Stokes $V/I_{c}$ [\%]} &\multicolumn{4}{c}{Maximum Stokes $Q/I_{c}$ [\%]}  \\
 \hline
&&TiO&MgH&CaH&FeH & TiO&MgH&CaH&FeH & TiO&MgH&CaH&FeH  \\
\hline
 M&3500K/2800K&  0.702&  0.671&  0.661&  0.688&  0.707&  0.758&  0.730&  0.704&  0.410&  0.153&   0.414&  0.166 \\
  &3500K/3000K&  0.705&  0.667&  0.660&  0.692&  0.988&  1.016&  1.078&  0.829&  0.529&  0.218&  0.599&  0.199 \\
  &3800K/3000K&  0.800&  0.690&  0.771&  0.723&  0.659&  0.602&  0.712&  0.623&  0.352&  0.125&  0.395&  0.150 \\
  &4000K/3000K&  0.866&  0.701&  0.830&  0.741&  0.498&  0.404&  0.535&  0.513&  0.266&  0.084&  0.297&  0.123 \\
  &4000K/3500K&  0.860&  0.692&  0.829&  0.746&  1.086&  1.058&  1.436&  0.877&  0.491&  0.226&  0.616&  0.222 \\
\hline 
K&4200K/3500K&   0.919&  0.715&  0.884&  0.774&  0.833&  0.724&  1.096&  0.729&  0.376&  0.155&  0.470&  0.184 \\
 &4500K/3500K&   0.970&  0.806&  0.957&  0.871&  0.600&  0.461&  0.784&  0.576&  0.270&  0.100&  0.336&  0.145 \\
 &4500K/3800K&   0.969&  0.800&  0.954&  0.871&  0.693&  0.790&  0.750&  0.790&  0.211&  0.170&  0.395&  0.204 \\
 &4800K/3800K&   0.981&  0.912&  0.978&  0.946&  0.534&  0.560&  0.575&  0.651&  0.162&  0.121&  0.303&  0.168 \\
 &4800K/4000K&   0.984&  0.906&  0.978&  0.945&  0.461&  0.816&  0.464&  0.798&  0.090&  0.170&  0.268&  0.221 \\
 &5000K/3800K&   0.984&  0.954&  0.983&  0.969&  0.459&  0.458&  0.492&  0.580&  0.139&  0.100&  0.260&  0.149 \\
 &5000K/4000K&   0.987&  0.948&  0.983&  0.968&  0.397&  0.670&  0.398&  0.712&  0.077&  0.140&  0.231&  0.197 \\
 &5000K/4200K&   0.992&  0.941&  0.985&  0.967&  0.248&  0.912&  0.283&  0.804&  0.040&  0.176&  0.168&  0.237 \\
\hline
G&5500K/3800K&   0.989&  0.987&  0.989&  0.987&  0.334&  0.302&  0.356&  0.456&  0.101&  0.066&  0.189&  0.117 \\
 &5500K/4200K&   0.994&  0.978&  0.990&  0.985&  0.182&  0.608&  0.206&  0.634&  0.030&  0.118&  0.123&  0.187 \\
 &5500K/4500K&   0.999&  0.977&  0.996&  0.989&  0.035&  0.580&  0.085&  0.471&  0.010&  0.182&  0.041&  0.149 \\
 &5800K/3800K&   0.990&  0.992&  0.991&  0.989&  0.285&  0.245&  0.302&  0.403&  0.086&  0.054&  0.161&  0.104 \\
 &5800K/4200K&   0.995&  0.984&  0.991&  0.987&  0.156&  0.494&  0.175&  0.562&  0.026&  0.096&  0.105&  0.166 \\
 &5800K/4500K&   0.999&  0.983&  0.997&  0.992&  0.030&  0.473&  0.073&  0.419&  0.009&  0.148&  0.034&  0.133 \\
 &6000K/3800K&   0.991&  0.993&  0.991&  0.990&  0.258&  0.215&  0.273&  0.374&  0.077&  0.048&  0.145&  0.096 \\
 &6000K/4200K&   0.995&  0.986&  0.992&  0.989&  0.141&  0.435&  0.159&  0.522&  0.023&  0.085&  0.095&  0.154 \\
 &6000K/4500K&   0.999&  0.986&  0.997&  0.992&  0.028&  0.418&  0.067&  0.389&  0.008&  0.130&  0.031&  0.123 \\
 &6000K/4800K&   1.000&  0.992&  0.999&  0.997&  0.003&  0.208&  0.017&  0.174&  0.002&  0.081&  0.008&  0.057 \\
\end{tabular}
\end{table*}

\noindent Magnetic fields play a crucial role in the activity phenomena on the Sun and on cool stars. Solar activity features have been studied in detail to reveal the structure of the underlying magnetic fields, which was used to constrain the solar dynamo theory. The detailed physics of stellar magnetic interactions, however, remains vague because we have insufficient direct measurements of magnetic fields on the surfaces of cool stars. A comprehensive overview of starspot phenomena on different types of cool stars, observational tools, and diagnostic techniques for studying starspots and their properties, such as magnetic fields, is given by \cite{berdyugina2005}. Stellar magnetic fields have been measured in various stellar objects with different techniques \citep[e.g.,~][]{valentikrull2001, donatilandstreet2009, reiners2012}.
Such  measurements can be achieved directly by the use of polarimetric techniques. The detection of magnetic fields on cool stars via polarimetry, however, is impeded on one hand by limited observational tools, that is, there are hardly any large telescopes with high-resolution spectropolarimeters. On the other hand, the net circular polarization signal from regions of mixed polarity fields of the unresolved stellar disk cancels out, which only allows detections of large-scale magnetic fields. The main part of our knowledge about magnetic fields on cool stars and in starspots is based on the analysis of Zeeman-broadened spectral lines. Zeeman signatures in atomic lines that are due to magnetic fields in starspots are diluted by contributions from the rest of the star. To detect such magnetic fields, the use of molecular over atomic lines is favorable because the latter receive a strong contribution from outside the spot umbra, whereas some molecular lines can only be formed in cool starspots if the effective temperature of the stellar photosphere is high enough. The choice of the adequate molecule therefore represents the key to measuring magnetic fields unambiguously in unresolved spots wherein the molecules are formed. Here, we investigate the four most prominent molecular bands of TiO, CaH, MgH, and FeH.

\noindent In contrast to sunspots, our knowledge of starspots and particularly of their magnetic properties is very scarce. However, over the past years, observational tools and diagnostic techniques for studying starspots have continued to improve. Long-term measurements (light-curve modeling) investigate the changes and periodicity of large-scale fields in stellar atmospheres. Mapping the starspot distribution on the stellar surface was made possible with the Doppler imaging technique, which uses high-resolution spectral line profiles of rapidly rotating stars. To reveal the magnetic field distribution in addition to the
temperature and abundance structure on the stellar surface, the magnetic Zeeman Doppler imaging method, based on the analysis of high-resolution spectropolarimetric data, is the appropriate tool \citep{semel1989, morinetal2013}. Recently, exoplanet transits, (spectro-) interferometry, microlensing, astereoseismology, etc. enlarged the diagnostic repertoire that is available for studying starspots.

The importance of starspots, in addition to studying the underlying magnetic field and its evolution, lies in their significance for the ongoing exoplanetary research. A planet transiting a starspot can reveal parameters of the exoplanetary system such as its obliquity, as shown in \citet{sanchisojedaetal2013} for a number of Kepler planet hosts. The presence of starspots outside the path of the transit is also relevant because it affects the depth of the transit or secondary eclipse from which the size and the albedo of the exoplanet can be deduced. To assess this effect, parallel observations of the transit or secondary eclipse curves as well as of the stellar activity are required. For the latter,  the properties of the starspot - such as magnetic field, temperature, and size - need to be monitored, and their effects on the observed signal need to be determined.

The technique applied throughout this work is the use of molecular lines as direct diagnostics of cool spots on the surfaces of active stars. Given a high photospheric temperature, particular molecular lines can only be formed in cool starspots, which allows detecting spots irrespective of their distribution, and furthermore, on slowly rotating stars.

\citet{vogt1979} reported on the first detection of molecular lines from starspots when he found TiO and VO bands on a K2 star, indicating the existence of a spot with a spectral type of an M6 star, where the molecular lines originated. \citet{huenemoerderramsey1987} and later \citet{neffetal1995} and \citet{onealetal1996} developed a technique for determining spot temperatures and filling factors from molecular lines. The method fits an observed spectrum with the sum of two components containing  effective temperatures of the spot and the unspotted photosphere weighted by spot filling factor and continuum surface flux ratio. Extending this basic approach to the polarimetric domain, we model the spectrum of a spotted star using G, K, and M type stars and apply various magnetic field strengths to estimate the expected (circular and linear) polarization signal. Considering the new generation of high-resolution spectropolarimeters, it is important to know which molecule is the most magnetically sensitive and produces a detectable signal in circular polarization. \citet{berdyuginaetal2006b} for the first time detected circular polarization of 0.5-1\% in TiO, CaH, and FeH lines on three active M dwarfs.

Detection of Zeeman-broadened or even Zeeman-split lines led to detections and estimates of magnetic fields on G, K, and M main-sequence stars.  As discussed in the comprehensive overview on starspots by \citet{berdyugina2005}, approximate values of starspot filling factors, temperatures, and magnetic field strengths are known. In particular, the models applied in this paper are based on observed combinations of spot and photosphere temperatures. 

For stellar surfaces, high filling factors of up to 50\% or even more have been determined by \citet{onealetal1996}. This unusual starspot size might be a composition of multiple sunspot-sized spots \citep{solanki2002}. \citet{berdyugina2002} discussed the reasons for the contradiction of such large spot filling factors with Doppler imaging results.

In contrast to the Sun, where large spot areas result in a higher luminosity due to a corresponding increase in plage area, the photometric and spectroscopic variability of active stars is dominated by the spot umbra \citep{radicketal1990}.
Low spot filling factors in G dwarfs were found. A representative sample of starspot temperatures on very active stars shows a clear tendency for spots to show a higher contrast to the photosphere than in hotter stars: the temperature difference between spots and the photosphere decreases from about 2000~K in G0 stars to 200~K in M4 stars, which holds for dwarfs and giants \citep{berdyugina2005}.

Cooler dwarfs have stronger magnetic fields that cover larger areas \citep{krullvalenti1996}. To detect magnetic fields in the starspot umbra, it is necessary to keep track of the spectral lines that are weak outside the spot and strong in the umbra; the goal of this work is to reveal such molecular lines.

In this analysis we investigate the temperature range in which the molecules MgH, TiO, CaH, and FeH can serve as indicators for magnetic fields on very active cool stars, and we present synthetic Stokes profiles for the modeled spectral type. This study is meant to serve as  basis for modeling and  planning spectropolarimetric observations.

\begin{figure*}[!ht]
\centering
\textbf{\large{Example G dwarf}}\par
\setlength{\unitlength}{1mm}
\begin{picture}(130,190)
\put(25,188){\textbf{MgH}}
\put(-25,80) {\begin{picture}(0,0) \includegraphics{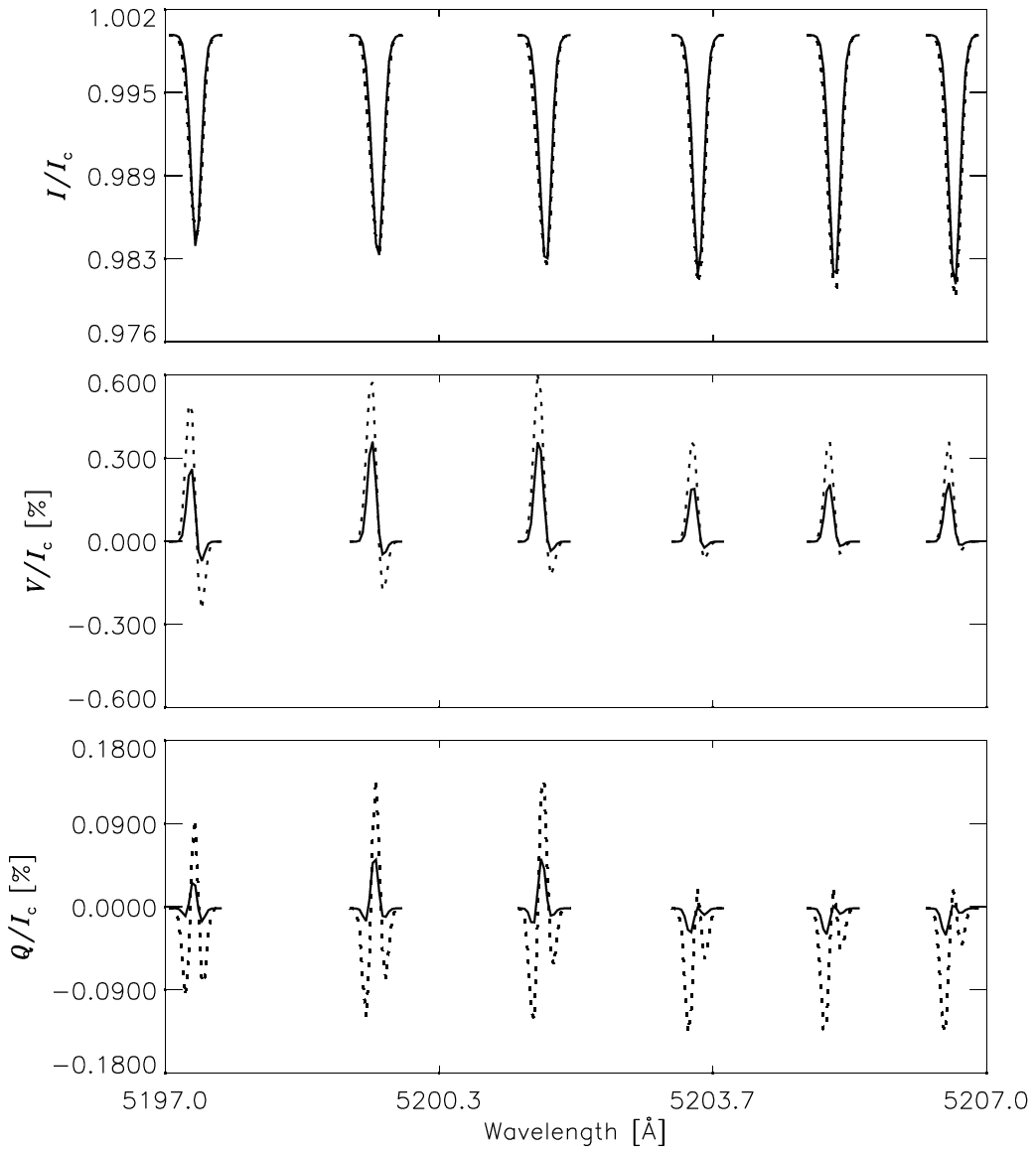} \end{picture}}
\put(105,188){\textbf{FeH}}
\put(55,80){\begin{picture}(0,0) \includegraphics{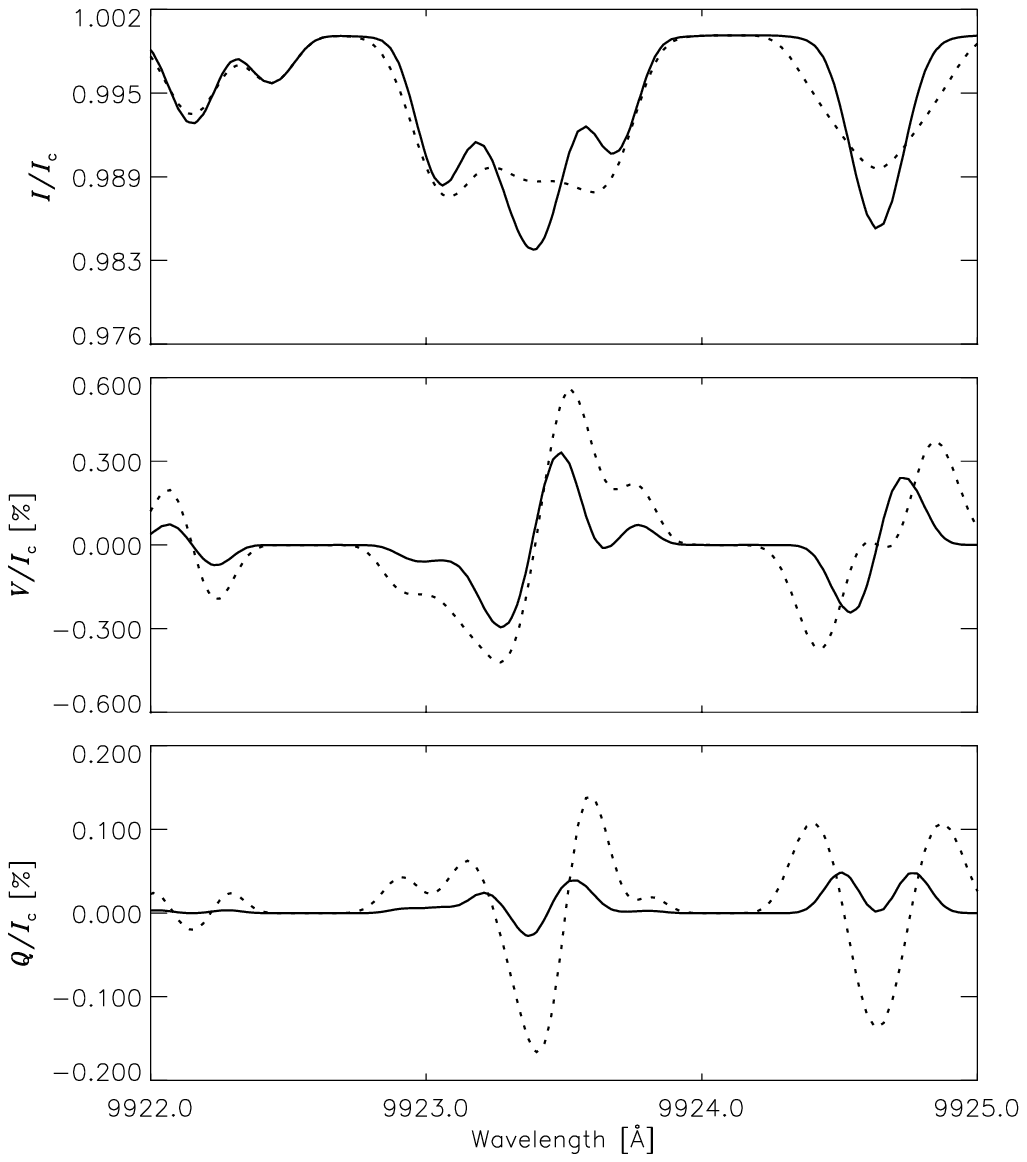} \end{picture}}
\put(25,94){\textbf{CaH}}
\put(-25,-17){\begin{picture}(0,0) \includegraphics{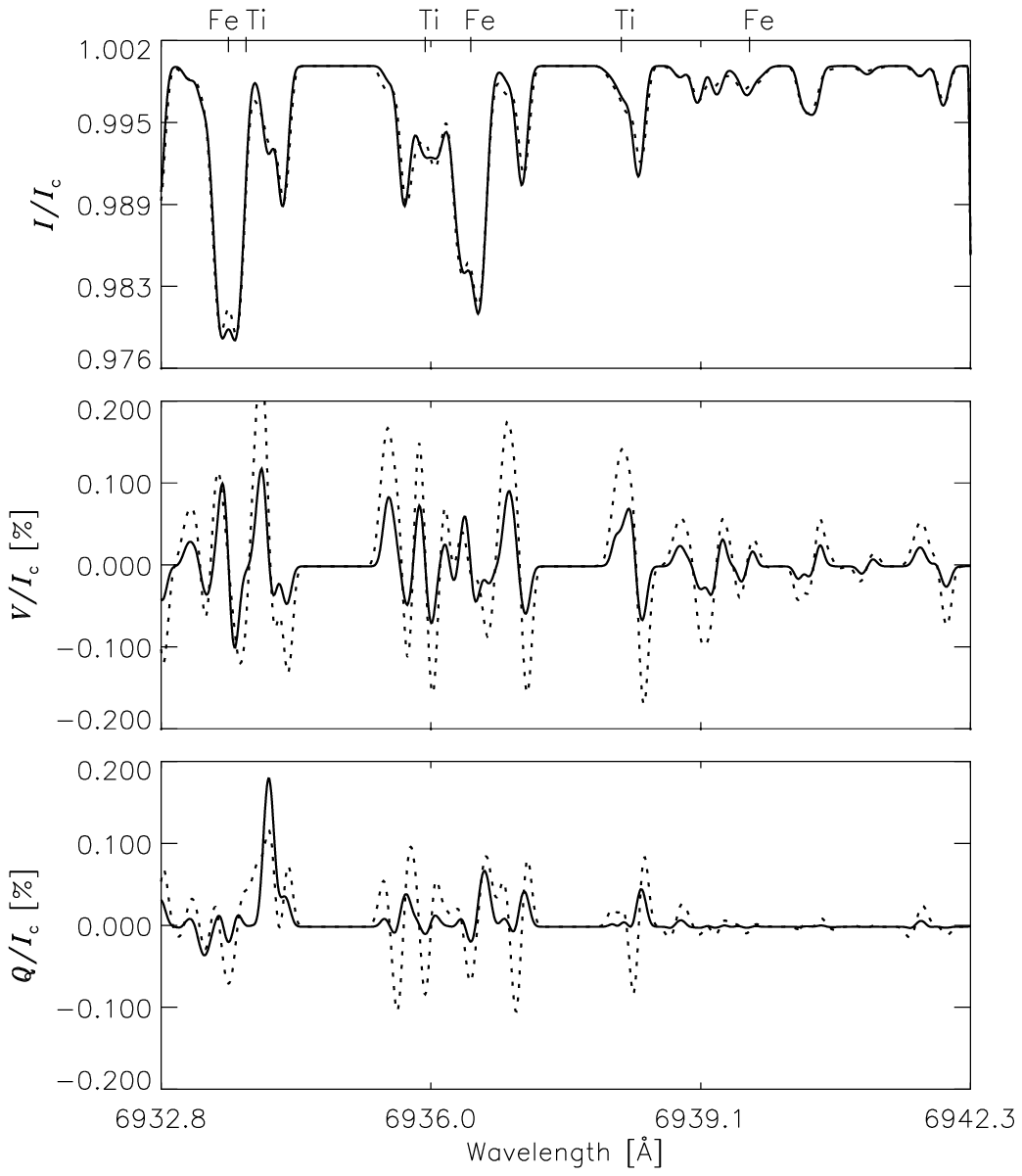} \end{picture}}
\put(105,94){\textbf{TiO}}
\put(55,-17){\begin{picture}(0,0) \includegraphics{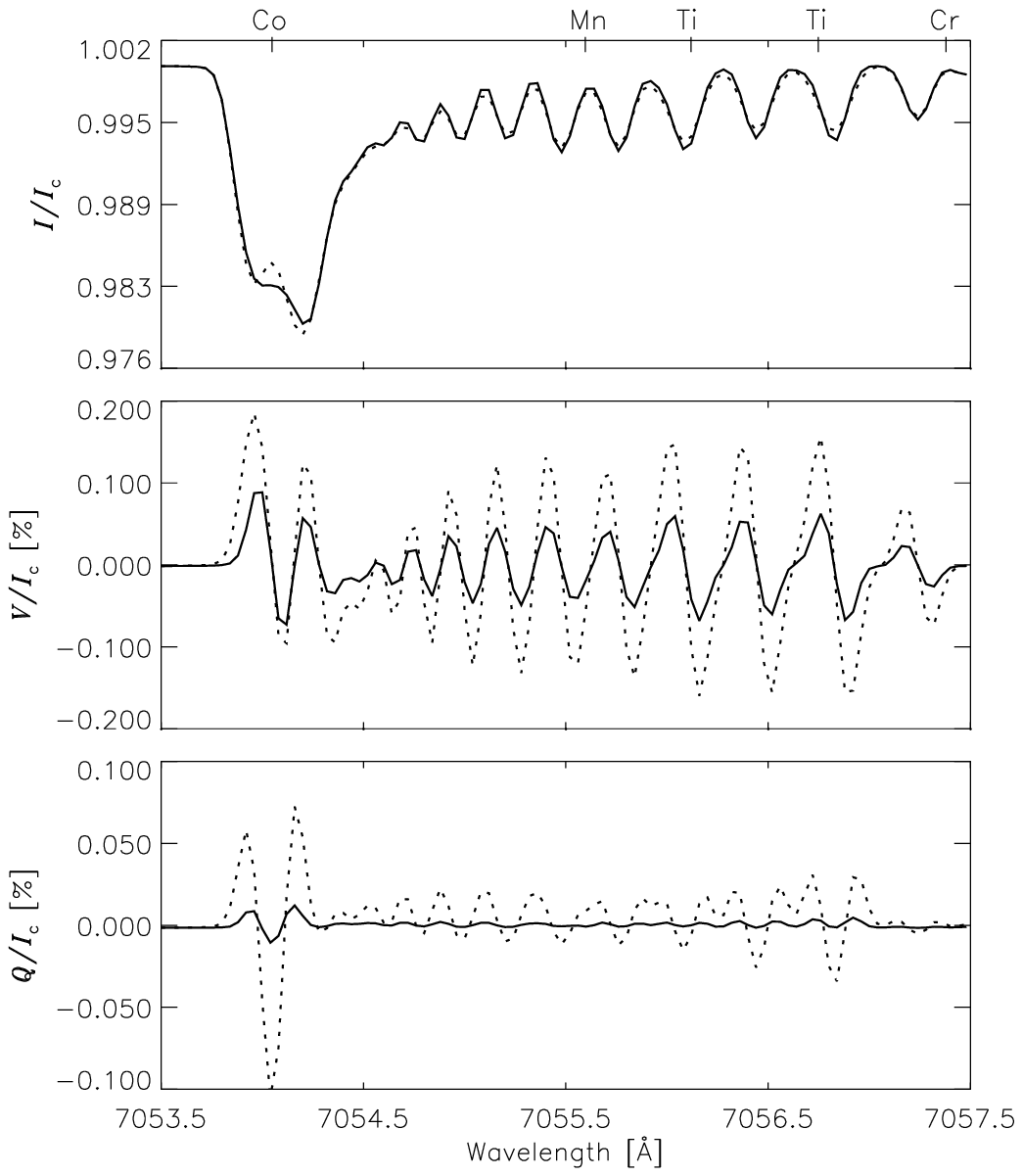} \end{picture}}
\end{picture}
\caption{Modeling of a G dwarf: Combinations of the photosphere and spot models for a model star with vsini=10~km/s, 10\% magnetic field of 1 kG (solid lines) and 3 kG (dashed lines), $T_{\rm phot}$=5800~K, $T_{\rm spot}$=4200~K.  Stokes V was calculated with an inclination angle of the magnetic field $\gamma=0^{\circ}$ and Stokes Q with $\gamma=90^{\circ}$.}
\label{fig:gstar}
\end{figure*}

\begin{figure*}[!ht]
\centering
\textbf{\large{Example K dwarf}}\par
\setlength{\unitlength}{1mm}
\begin{picture}(130,190)
\put(25,188){\textbf{MgH}}
\put(-25,80) {\begin{picture}(0,0) \includegraphics{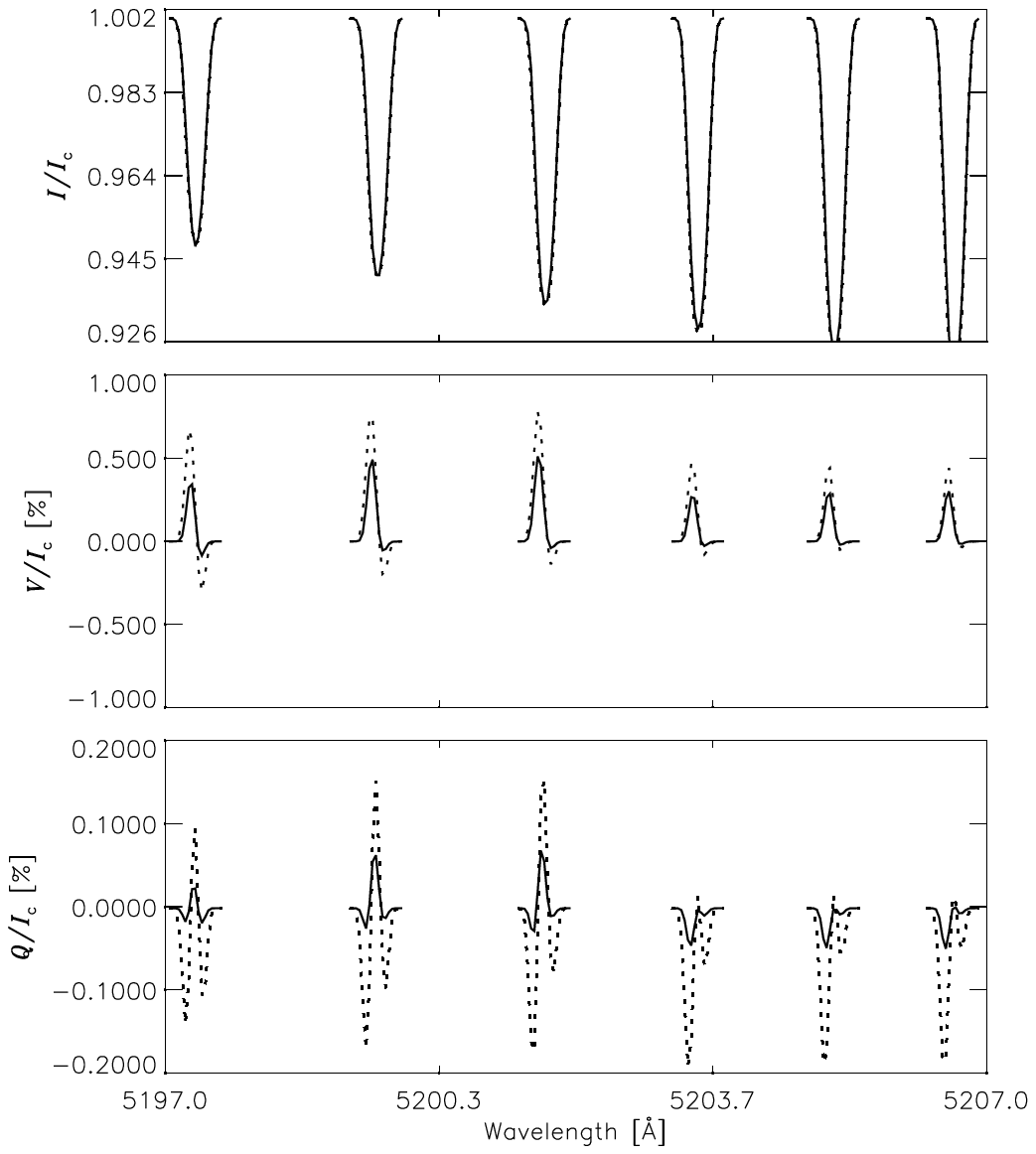} \end{picture}}
\put(105,188){\textbf{FeH}}
\put(55,80){\begin{picture}(0,0) \includegraphics{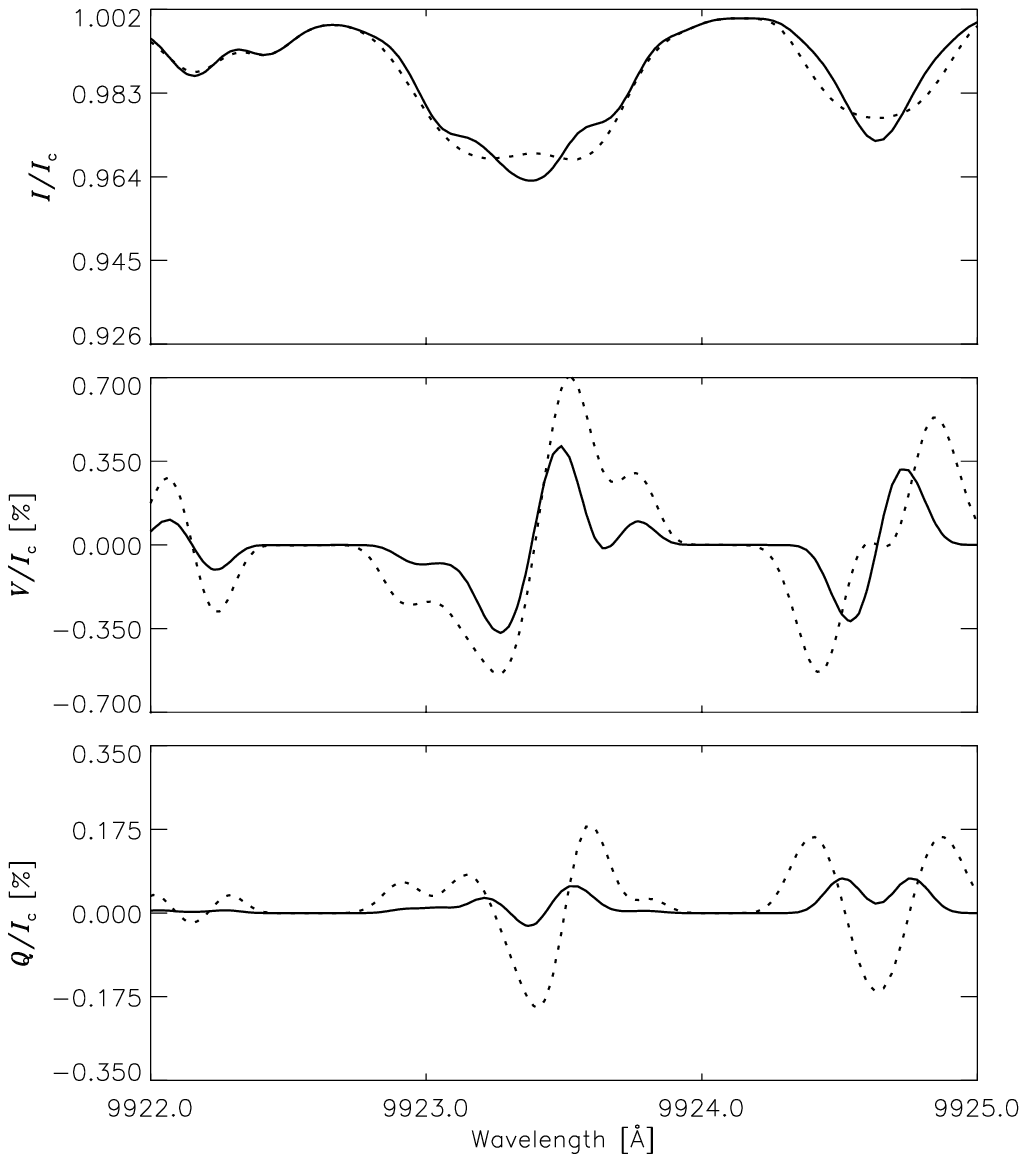} \end{picture}}
\put(25,94){\textbf{CaH}}
\put(-25,-17){\begin{picture}(0,0) \includegraphics{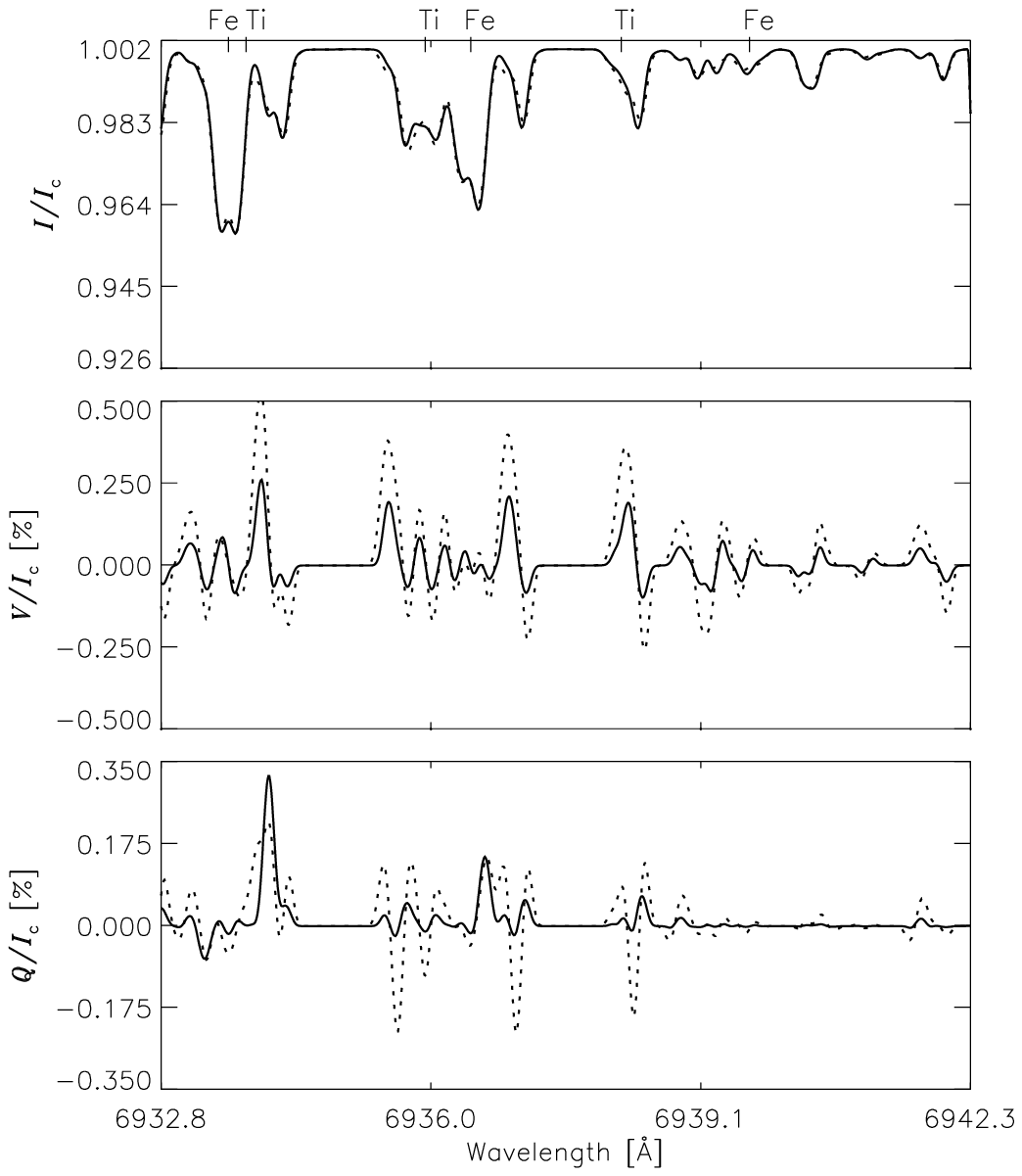} \end{picture}}
\put(105,94){\textbf{TiO}}
\put(55,-17){\begin{picture}(0,0) \includegraphics{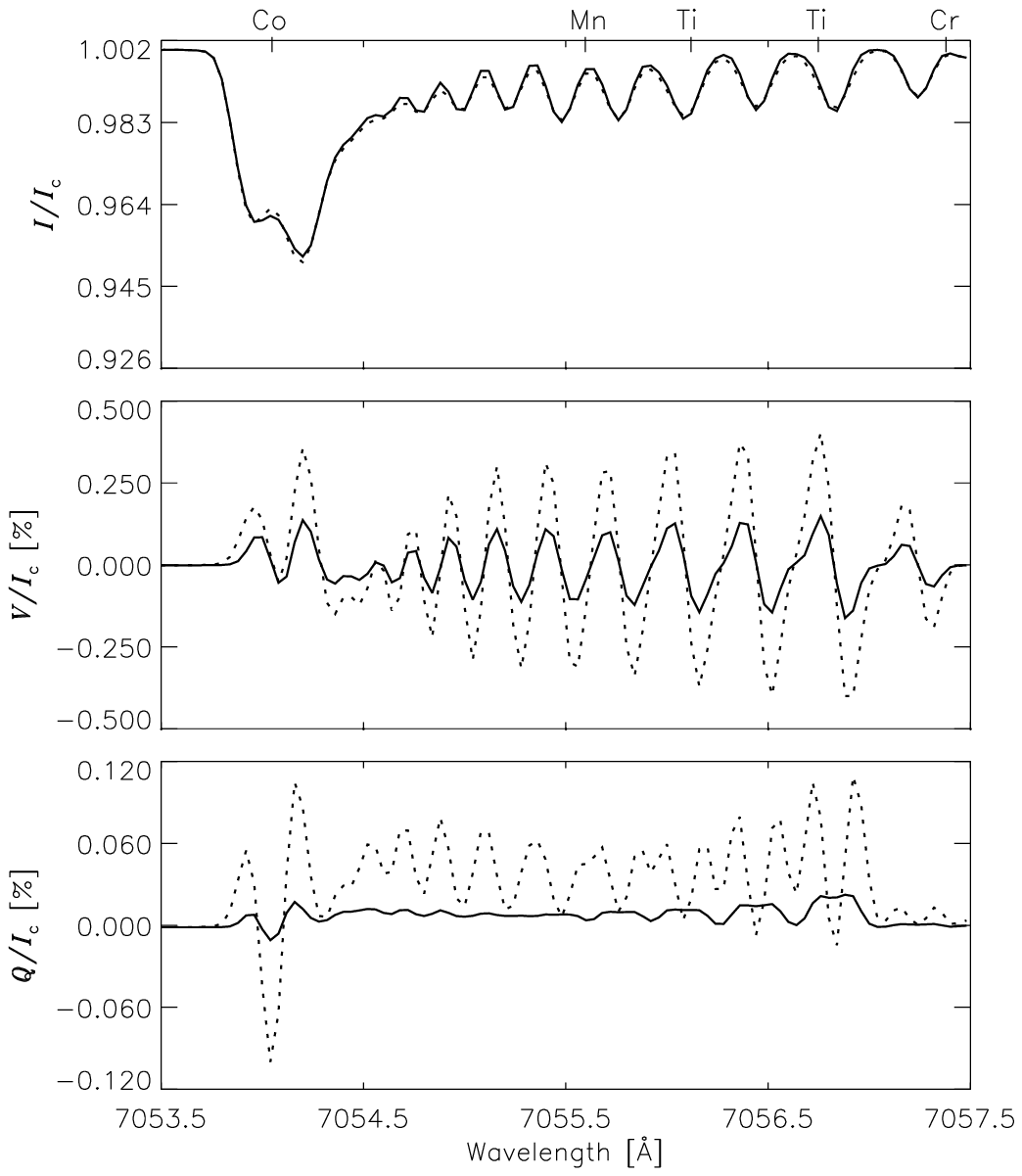} \end{picture}}
\end{picture}
\caption{Modeling of a K dwarf: Combinations of the photosphere and spot models for a model star with vsini=10~km/s, 10\% magnetic field of 1 kG (solid lines) and 3 kG (dashed lines), $T_{\rm phot}$=5000~K, $T_{\rm spot}$=4000~K. Stokes V was calculated with an inclination angle of the magnetic field $\gamma=0^{\circ}$ and Stokes Q with $\gamma=90^{\circ}$.}
\label{fig:kstar}
\end{figure*}

\begin{figure*}[!ht]
\centering
\textbf{\large{Example M dwarf}}\par
\setlength{\unitlength}{1mm}
\begin{picture}(130,190)
\put(25,188){\textbf{MgH}}
\put(-25,80) {\begin{picture}(0,0) \includegraphics{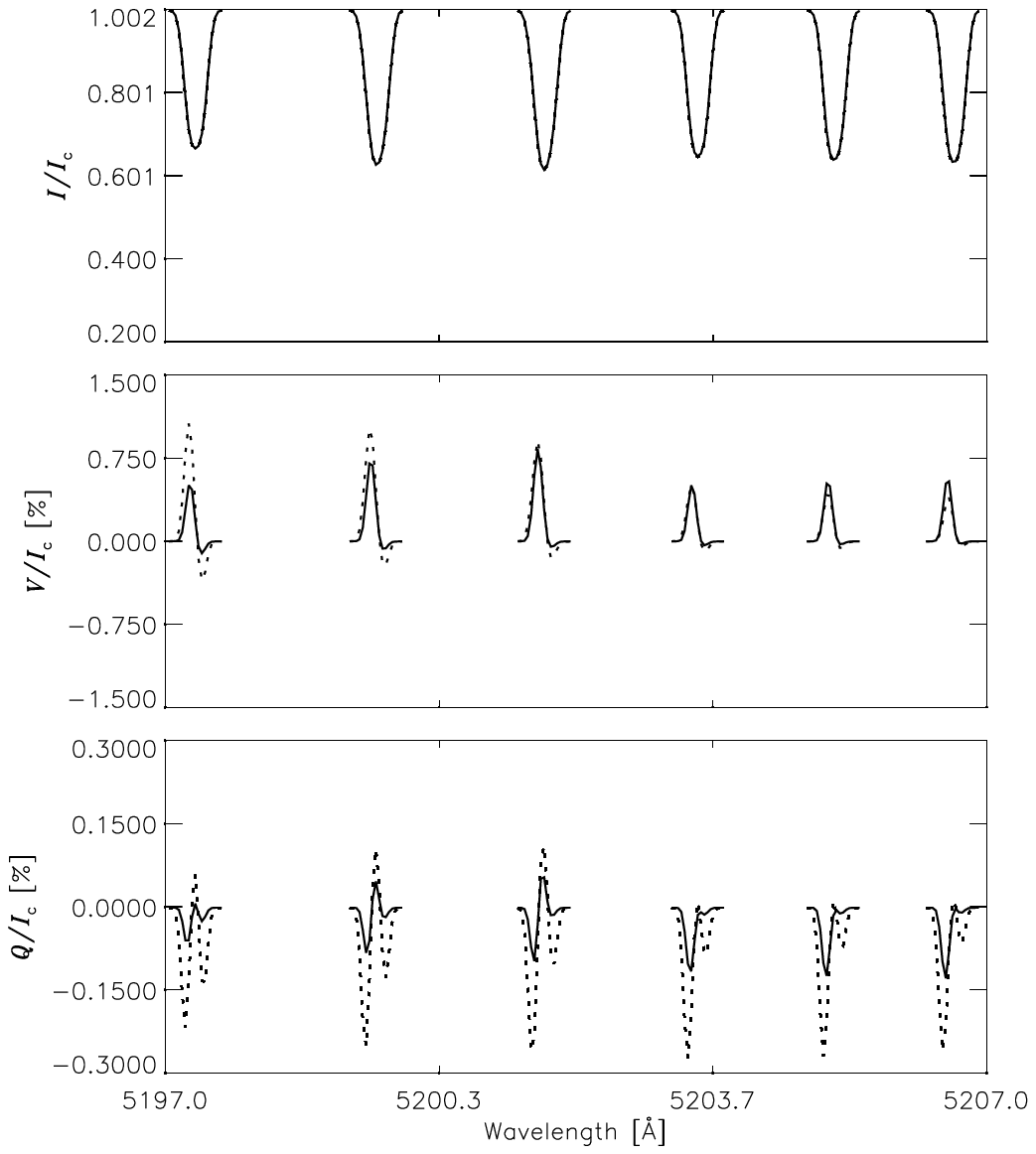} \end{picture}}
\put(105,188){\textbf{FeH}}
\put(55,80){\begin{picture}(0,0) \includegraphics{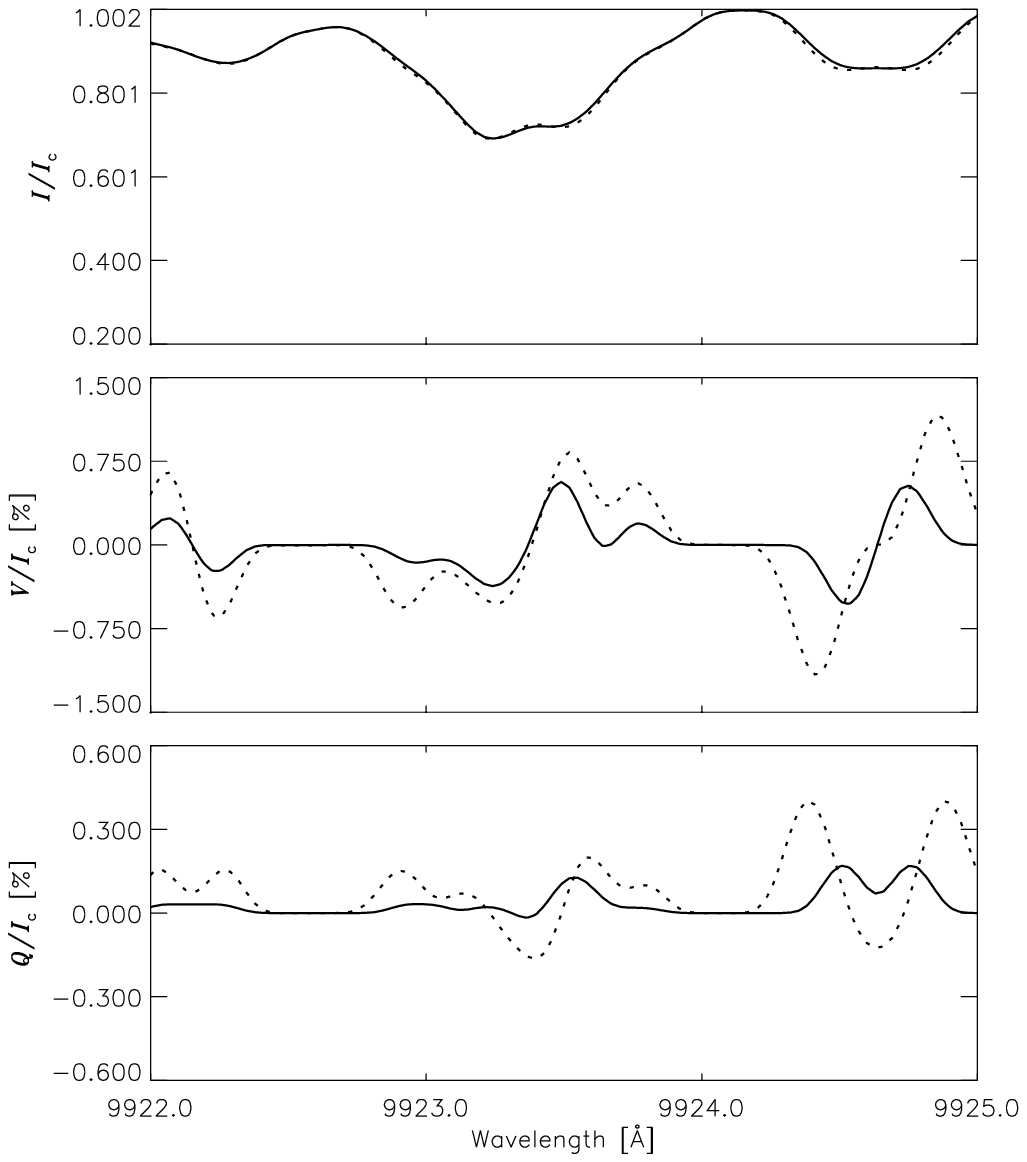} \end{picture}}
\put(25,94){\textbf{CaH}}
\put(-25,-17){\begin{picture}(0,0) \includegraphics{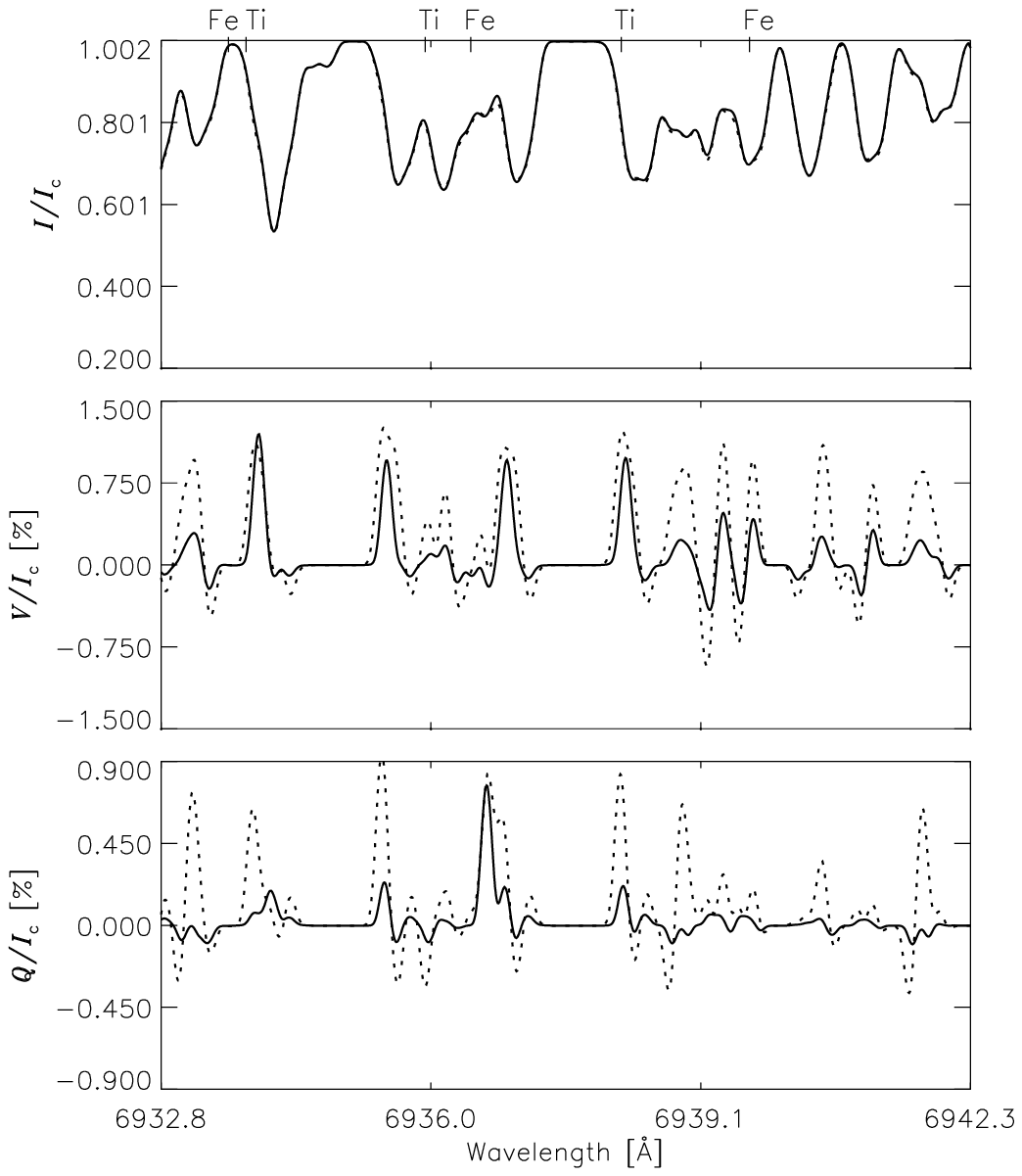} \end{picture}}
\put(105,94){\textbf{TiO}}
\put(55,-17){\begin{picture}(0,0) \includegraphics{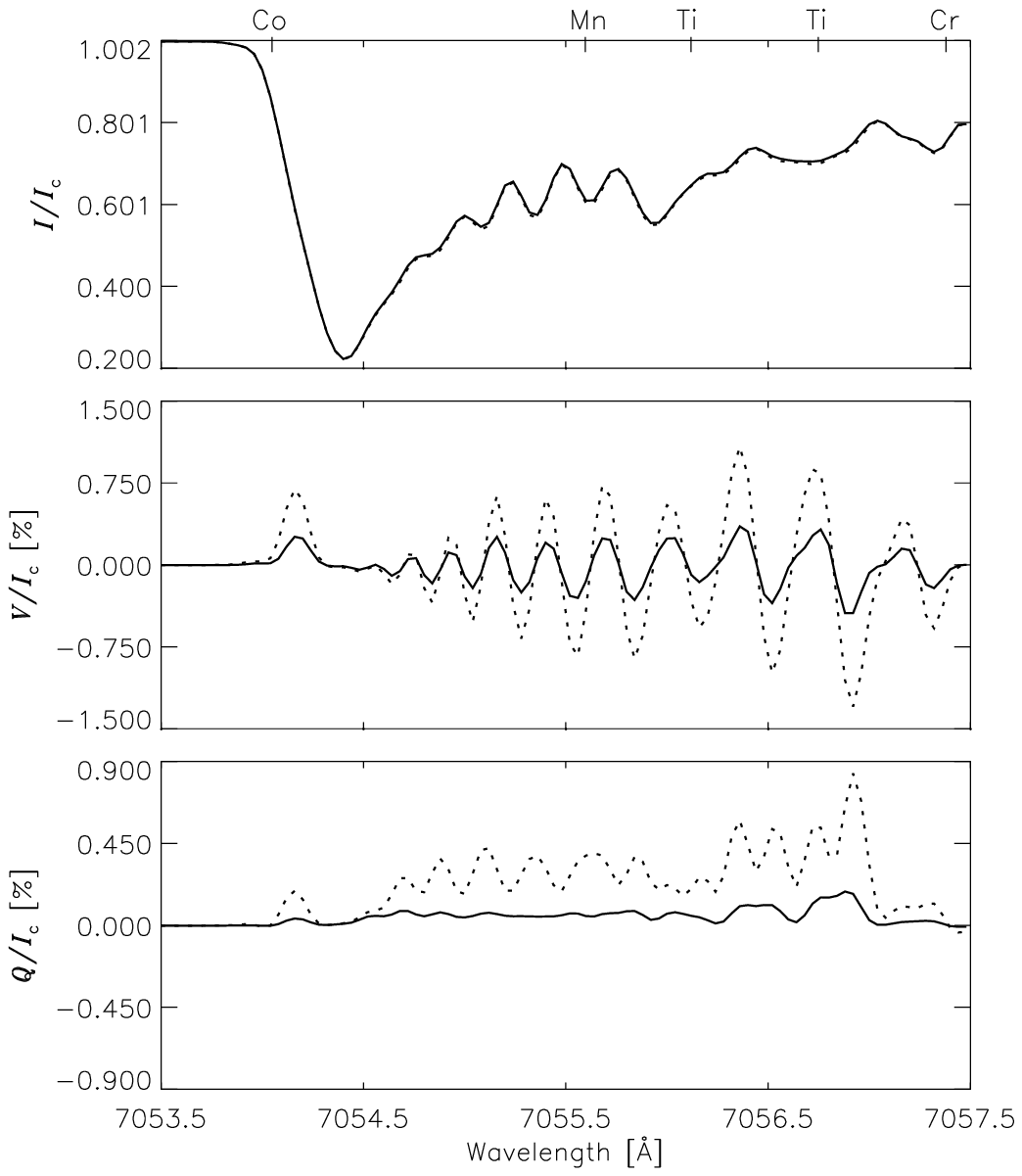} \end{picture}}
\end{picture}
\caption{Modeling of an M dwarf: Combinations of the photosphere and spot models for a model star with vsini=10~km/s, 10\% magnetic field of 1 kG (solid lines) and 3 kG (dashed lines), $T_{\rm phot}$=3500~K, $T_{\rm spot}$=3000~K. Stokes V was calculated with an inclination angle of the magnetic field $\gamma=0^{\circ}$ and Stokes Q with $\gamma=90^{\circ}$.}
\label{fig:mstar}
\end{figure*}

\section{Molecular bands}\label{diag}

TiO lines are prominent in spectra of sunspots and cool stars. TiO was first astronomically detected in spectra of M dwarfs \citep{fowler1904}. The first detection of TiO molecular bands from starspots was reported by \citet{vogt1979} for a K2 star. \citet{ramseynations1980} suggested using TiO bands to measure starspot properties. Further observations of TiO bands in spectra of active stars allowed the spot area and temperature to be measured \citep[e.g.,~][]{neffetal1995, onealetal1996}. \citet{valentietal1998} explored the extent to which the spectral synthesis of TiO lines could reproduce the optical spectrum of an inactive M dwarf. For sunspots, the first spectropolarimetric models and observations were shown in \citet{berdyuginaetal2000}. \citet{berdyuginaetal2003} concluded that the TiO $\gamma(0,0)R_3$ system band head at 7054\,\AA\ is one of the best molecular diagnostics of the magnetic field in (sunspot umbra and) starspots. They expected a Stokes V signal of 0.3\% and were able to nicely reproduce TiO line polarization observed on three M dwarfs  \citep{berdyuginaetal2006b}. The triplet states of this system can be described sufficiently well by Hund's case (a), especially for transitions between lower rotational levels.    \citet{onealetal2004} presented measurements of starspot temperature and filling factor on five highly active stars, using absorption bands of TiO. They fitted TiO bands using spectra of inactive G and K stars to represent the unspotted photospheres of the active stars and spectra of M stars to represent the spots.

For MgH (and other diatomic molecules), \citet{kronig1928} and \citet{hill1929} theoretically predicted the molecular Zeeman splitting. Laboratory measurements of the Zeeman effect were reported in \citet{crawford1934}. MgH spectral lines were described in \citet{laborde1961}, where a comparison of laboratory wavelengths with solar ones was given. \citet{belletal1985} used MgH lines to determine the surface gravity of Arcturus. \citet{berdyuginasavanov1992} modeled the MgH radical as a diagnostic of the surface gravity of K giants. \citet{berdyuginaetal2000} presented the first spectropolarimetric measurements and modeling of MgH lines in sunspots. The A$^{2}\Pi$ state of the MgH A$^{2}\Pi$--X$^{2}\Sigma^{+}$ (0,0) transition is an intermediate Hund case (a-b) with spin-orbit constants of $Y =5.7$, meaning that this system is relatively close to a pure Hund case (b). With the Paschen-Back effect taken properly into account, the MgH A$^{2}\Pi$--X$^{2}\Sigma^{+}$ system at 5200\,\AA\  represents a sensitive tool  for stellar magnetic studies \citep{berdyuginaetal2000, berdyuginaetal2005}.

Because it is one of the most important astrophysical molecules, CaH bands can also be used as a luminosity indicator of cool stars \citep{oehman1934, mould1976, mouldwallis1977, barbuyetal1993}. CaH bands represent a valuable tool for studying brown dwarfs since the absorption therein is an important opacity source. The CaH A$^{2}\Pi$--X$^{2}\Sigma^{+}$ band system is observed in the wavelength region at 6600--7600\,\AA. The ground state X$^{2}\Sigma^{+}$ is well described by a pure Hund case (b). The excited state A$^{2}\Pi$ is intermediate between Hund cases (a) and (b). Like for MgH, the Paschen-Back effect must be considered \citep{berdyuginaetal2003}. The first polarimetric measurements and modeling of CaH Stokes profiles was carried out by \citet{berdyuginaetal2006a}. They confirmed the diagnostic value of CaH  in cool astrophysical sources.

\citet{valentietal2001} clearly showed the usefulness of the FeH lines as stellar magnetic diagnostics with their detection of Zeeman broadening of FeH lines in an active M dwarf. They also modeled the stellar spectrum using the sunspot spectrum by \citet{wallaceetal1998}. In these sunspot observations, a remarkable magnetic sensitivity of many lines of the FeH F$^{4}\Delta$--X$^{4}\Delta$ system can be seen. It was concluded that this particular molecular system would be a powerful tool for diagnosing solar and stellar magnetism once the spin-coupling constants became available \citep[e.g.,~][]{berdyuginaetal2000}. \citet{reinersbasri2006} measured magnetic fields on ultracool stars using FeH and emphasized the fact that its response to magnetic fields is easier to detect than other magnetic diagnostics.  The FeH F$^{4}\Delta$--X$^{4}\Delta$ system is produced by transitions between two electronic states with couplings of the angular momenta that are intermediate between limiting Hund cases (a) and (b) \citep{phillipsetal1987}. Its high magnetic sensitivity was pointed out by \citet{wallaceetal1998}. With the help of the previously missing spin-coupling constants provided by \citet{dulicketal2003}, the perturbation calculation of the molecular Zeeman effect for this transition allowed \citet{aframetal2007, aframetal2008} to analyze the FeH F$^{4}\Delta$--X$^{4}\Delta$ system and its diagnostic capabilities for investigating solar and stellar magnetic fields. They presented fits to the first spectropolarimetric measurements in sunspots \citep{aframetal2007}, providing Land\'e factors for energy levels and transitions. They also compared synthetic and observed Stokes profiles in different wavelength regions \citep{aframetal2008}.

\subsection{The model}\label{model}
\noindent The Stokes parameters $I$, $V$, and $Q$ were calculated with the radiative transfer code STOPRO \citep{solanki1987, berdyuginaetal2003},
which assumes local thermodynamic equilibrium and solves the radiative transfer equations upon input of molecular or atomic linelists and an atmospheric model.
For this work we used Phoenix stellar models \citep{hauschildtetal1999} in local thermodynamic equilibrium (LTE) with an effective temperature range of 2800~K to 6000~K and a surface gravity of 4.5, corresponding to G, K, and M dwarfs.
In dwarfs, the diatomic hydrides MgH, FeH, and CaH are more abundant and hence stronger than in giants \citep{belletal1985, barbuyetal1993}. However, the magnetic signatures studied in this work are independent of the log g value. In contrast, other diatomic molecules, such as CN and CO, are stronger in giants. 
 The radiative transfer code STOPRO treats the spectral line in a  magnetized atmosphere accounting for the atomic and molecular Zeeman and Paschen-Back effects.  
If not declared otherwise, we assumed the star to have a circular spot with an area of 10\% of the stellar disk and a spot comprised of only longitudinal  or transverse magnetic fields. We obtained the spectrum of the spotted star by integrating synthetic polarized spectra over the stellar surface taking into account limb darkening, stellar rotation, and the direction of the magnetic field. We estimated the strength of the circular and linear polarization Stokes $V$ and $Q$ signals for the different molecules MgH, TiO, CaH, and FeH. We combined various photosphere and spot models according to the current knowledge of stellar properties \citep{berdyugina2005}. For CaH and TiO, a number of atomic lines are included in the calculations, as indicated in Figs.~\ref{fig:gstar}-\ref{fig:mstar}. We took into account the following line-broadening effects: intrinsic, thermal (according to the local temperature in the model atmosphere), magnetic, rotational (10~km/s), microturbulent (1~km/s), and instrumental (0.07\,\AA) broadening.

\section{Modeled signals for G-K-M dwarfs}\label{signals}

\subsection{G dwarfs}\label{gstars}

\noindent Type G dwarfs with photosphere temperatures of 5000~K-- 6000~K show an average spot temperature $T_{\rm spot}$ of 70\% of $T_{\rm phot}$. There are exceptions, however, such as EK Dra, which have a lower contrast between $T_{\rm phot}$ and $T_{\rm spot}$ ($T_{\rm phot}=5900~K$, $T_{\rm spot}=4900~K$). Spot filling factors are relatively low, ranging from 10\% -- 25\% on active dwarfs and from 30\% -- 35\% on active giants. The Sun has a spot-filling factor of 1\%.
Magnetic field measurements for the Sun show an umbral magnetic field strength of 3 kG.

In Fig.~\ref{fig:gstar} we present synthetic Stokes profiles for the molecules we investigated for a G dwarf. For clarity, a single type of a G star is presented in the figure, with the combination of the photosphere model with $T_{\rm phot}$=5800~K and a spot model $T_{\rm spot}$=4200~K (see Table \ref{tab:overview} for all calculated combinations). We applied an instrumental broadening of 0.07\,\AA\ and a rotational broadening corresponding to v~sin~i=10~km/s.

In each figure, the middle panel (Stokes V) was calculated with an inclination angle $\gamma=0^{\circ}$ (longitudinal magnetic field) and the lowest panel with an inclination angle $\gamma=90^{\circ}$ transverse magnetic field). In the latter case,
a relatively strong Stokes Q can be seen.
Strong asymmetries in the polarization signals for MgH and CaH occur as a result of the Paschen-Back effect and lead to a net polarization signal across the line profiles. In Fig.~\ref{fig:gstar}, for TiO some strong polarization signals are due to blended atomic lines, for example, the Co line near 7054\,\AA. Table \ref{tab:overview} tabulates the maximum amplitude of the polarization signal only from the molecular species, which forms exclusively in the spot.

Resulting minimum residual intensities, the maximum Stokes V and Q  amplitudes (in units of the continuum intensity) for MgH, FeH, CaH, and TiO are given in Table \ref{tab:overview} for all calculated G dwarfs.

\subsection{K dwarfs}\label{kstars}

\noindent The effective temperatures of K stars range from 4000~K to 5000~K.  Spot temperatures $T_{\rm spot}$ on K dwarfs average  73\% of $T_{\rm phot}$  \citep{berdyugina2005}.
Spot filling factors have values from 20\% -- 60\% and magnetic field filling factors range from 10\% -- 80\%, for typical magnetic field strengths of 1.5 kG -- 2.5 kG, with one exception of 5 kG.
In Fig.~\ref{fig:kstar}, results for a single type of K star are presented for the molecules MgH, FeH, CaH, and TiO, combining a photosphere model with $T_{\rm phot}$=5000~K and a spot model $T_{\rm spot}$=4000~K. For TiO, a broad Stokes Q signal can be seen, which is promising to be detected even with low-resolution observations.
See Table \ref{tab:overview} for the exact values of the residual intensity, Stokes V, and Stokes Q signals and for all other calculated K dwarf combinations.

\subsection{M dwarfs}\label{mstars}

\noindent The effective temperature for an M dwarf lies between 2600~K and 3800~K. Spot temperatures tend to be closer to $T_{\rm phot}$ , with $T_{\rm spot}$ being approximately 86\% of $T_{\rm phot}$.
Observed spot  filling factors are low with 5\% --20\%, but magnetic field filling factors show high values from 50\% -- 85\% (see discussion in Sect. \ref{sens}). The range of observed magnetic field strengths is rather wide with 2.5 kG -- 4.3 kG \citep{berdyugina2005}.
To model an M dwarf (Fig.~\ref{fig:mstar}),  we combined a photosphere model with $T_{\rm phot}$=3500~K and a spot model $T_{\rm spot}$=3000~K for the molecules MgH, FeH, CaH, and TiO. In the TiO spectrum, a broad Stokes Q signal is present here as well. In M dwarfs, atomic lines are much more blended with molecular lines than in G and K dwarfs. We tried to select wavelength regions with
a lower atomic line contribution than from molecular lines. We find that for MgH in M dwarfs, the strength of the Stokes V signal for some lines is almost the same, regardless  of the magnetic field, because of line saturation. In addition, the Paschen-Back effect causes nonlinear blending of the Zeeman components. Stokes Q is not saturated because it has more allowed Zeeman transitions, and, therefore, saturation does not occur yet.
We list all minimum residual intensities, Stokes V, and Stokes Q signals for all calculated M dwarf combinations in Table \ref{tab:overview}.

\subsection{Sensitivity of the modeled signal}\label{sens}
\begin{figure}
\centering
\textbf{\large{Sensitivity}}\par
\setlength{\unitlength}{1mm}
\begin{picture}(100,80)
\put(7,77){\textbf{spot size variation}}
\put(-97,-12) {\begin{picture}(0,0) \includegraphics{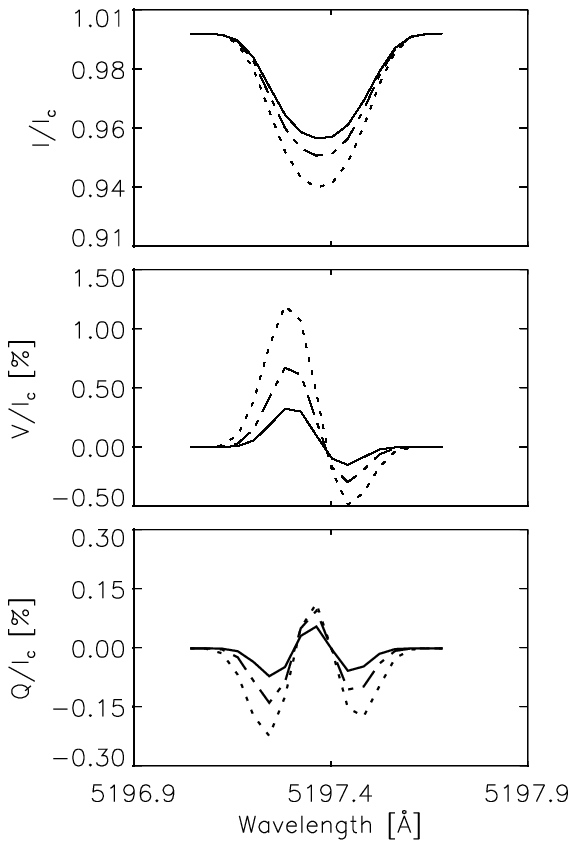} \end{picture}}
\put(49,77){\textbf{magnetic field variation}}
\put(-50,-12){\begin{picture}(0,0) \includegraphics{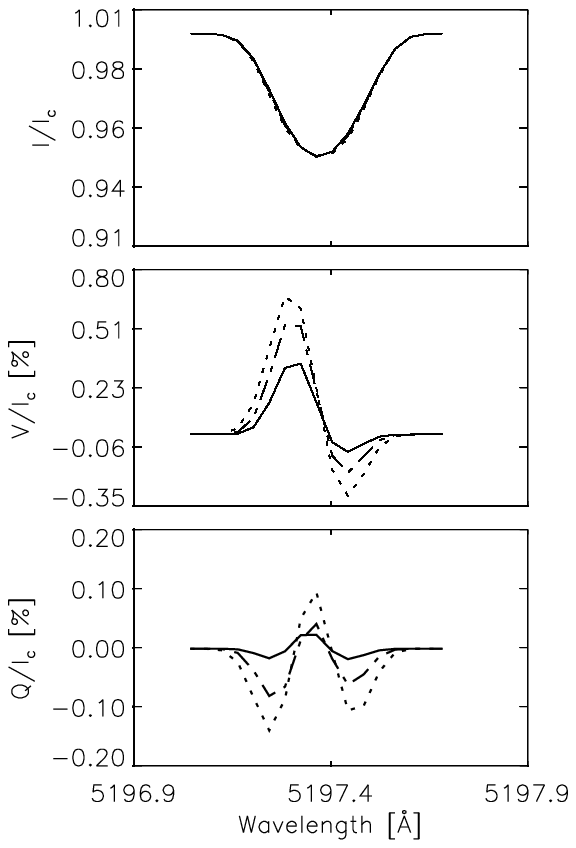} \end{picture}}
\end{picture}
\caption{Combination of the photosphere and spot models for the MgH A$^{2}\Pi$--X$^{2}\Sigma^{+}$ system at 5200\,\AA\ for a model star with vsini=10~km/s$, T_{\rm phot}$=5000~K,  $T_{\rm spot}$=4000~K, and with (i) left panel: 5\%, 10\%, and 20\% (solid, dashed, and dotted lines, respectively) magnetic field of 3~kG (ii) right panel: 10\% magnetic field of 1~kG, 2~kG, and 3~kG (solid, dashed, and dotted lines, respectively) magnetic field. Stokes V was calculated with an inclination angle of the magnetic field $\gamma=0^{\circ}$ and Stokes Q with $\gamma=90^{\circ}$.}
\label{fig:sensspot}
\end{figure}

To illustrate the sensitivity of the modeled signal to changes in spot properties, in Fig. \ref{fig:sensspot} we have modeled  a K dwarf for the MgH A$^{2}\Pi$--X$^{2}\Sigma^{+}$ system at 5200\,\AA\ with three values of the spot size: 5\%, 10\%, and 20\% (in the left panel) and with three values of the magnetic field strength:  1~kG, 2~kG, and 3~kG (right panel). In this regime of the Paschen-Back effect, the signal still increases with apparent spot size or magnetic field strength. These spot sizes can still be called moderate, given that filling factors of up to 80\% have been reported \citep{berdyugina2005}. In fact, the magnetic field can be complex, and opposite-sign-polarity fields cancel out in the polarization signal, so that the actual magnetic filling factor can be much higher than the assumed 10\%.
The minimum residual intensity reaches  0.96, 0.95, and 0.94, the Stokes V amplitudes (in units of the continuum intensity) are  0.32\%, 0.67\%, and 1.20\%, and the Stokes Q amplitudes (in units of the continuum intensity) are 0.07\%, 0.14\%, and 0.22\% for spot sizes of 5\%, 10\%, and 20\% respectively.
 For magnetic field strengths of 1~kG, 2~kG, and 3~kG, the minimum residual intensity is 0.95 in all cases, the Stokes V amplitudes (in units of the continuum intensity) are  0.34\%, 0.53\%, and 0.67\%, and the Stokes Q amplitudes (in units of the continuum intensity) are 0.02\%, 0.08\%, and 0.14\%.

\begin{figure*}[!ht]
\centering
\begin{picture}(130,220)
\put(-180,-30) {\begin{picture}(0,0) \includegraphics{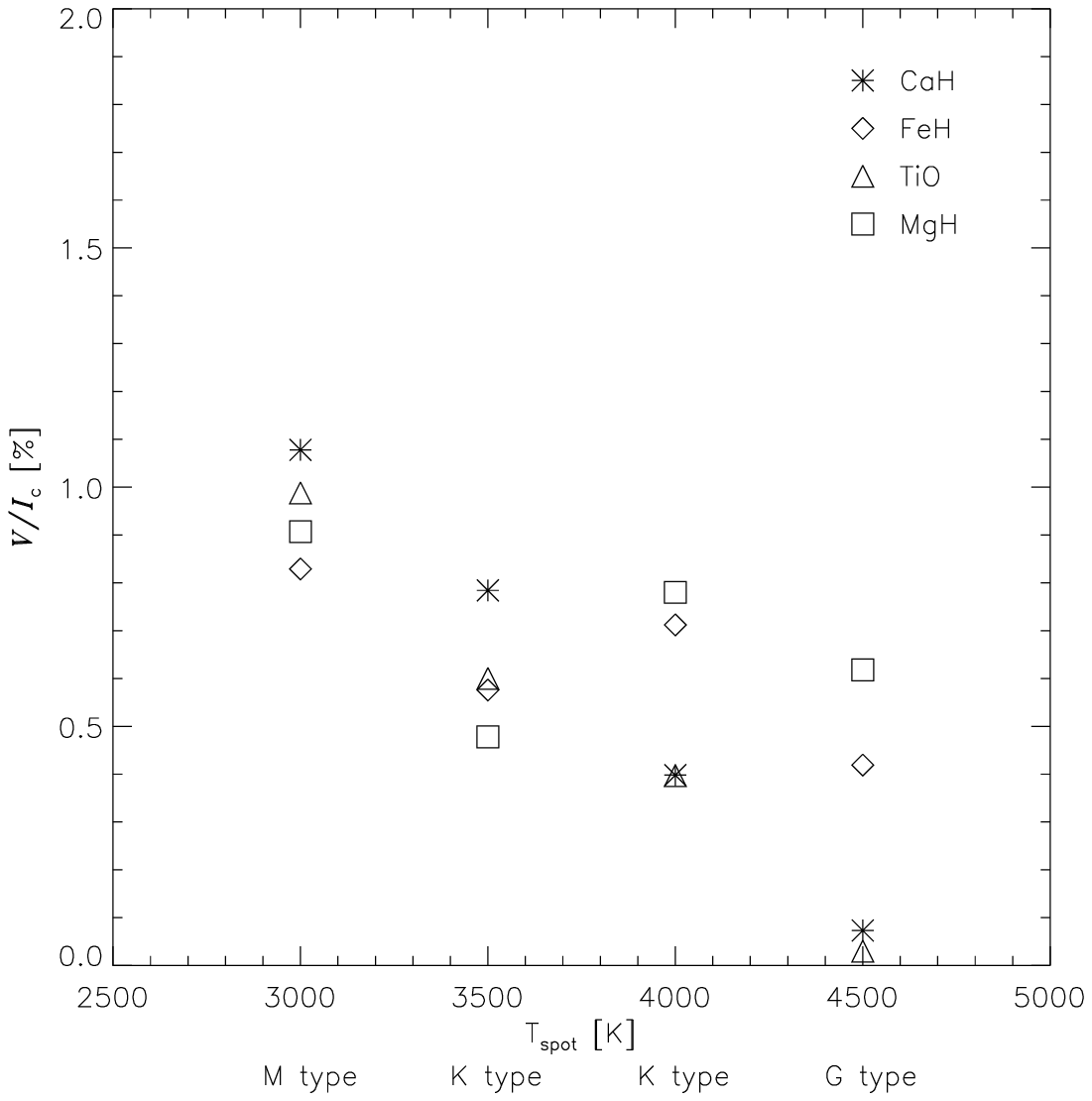} \end{picture}}
\put(50,-30) {\begin{picture}(0,0) \includegraphics{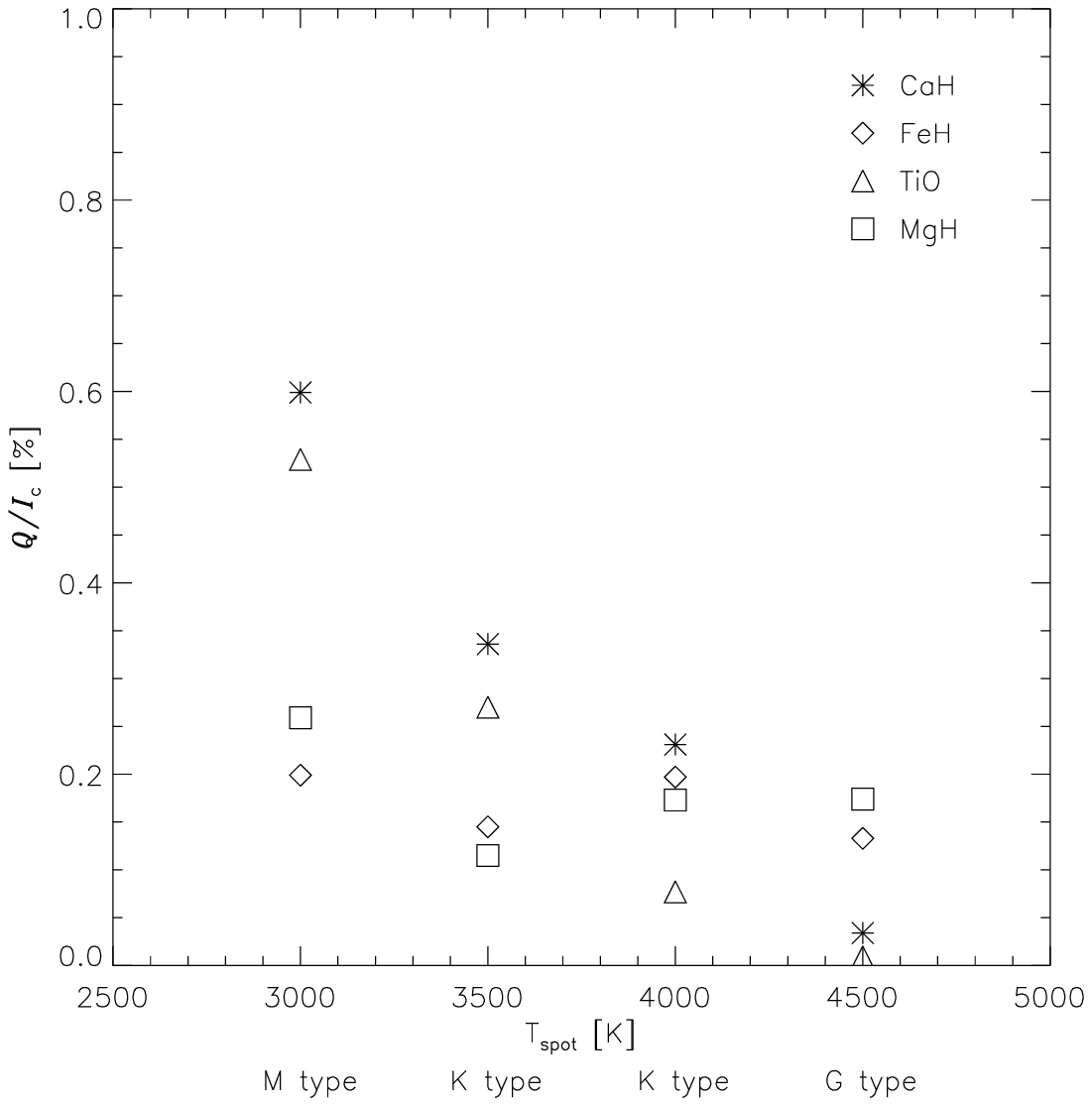} \end{picture}}
\end{picture}
\caption{Maximum Stokes $V$  signal (left panel) and $Q$  signal (right panel) from the different molecular lines for different spectral types for a model star with vsini=10~km/s and 10\% longitudinal or transverse magnetic field of 3 kG. In the plot, the applied spot temperatures are given and the photosphere temperatures are implied by the indicated spectral types.}
\label{fig:vq}
\end{figure*}

\section{Conclusions}
 We modeled various scenarios to calculate the expected circular and linear polarization signal in MgH, TiO, CaH, and FeH lines present in starspots. 
 We used a model assuming a spot size of 10\% of the stellar disk comprising either longitudinal or transverse magnetic fields and estimated the strength of the Stokes $V$ and $Q$ signals in molecular lines that could be expected to be observed from starspots on G-K-M dwarfs using realistic estimates of starspot parameters.  For M dwarfs and cooler K dwarfs, it should be noted that molecular lines can also be formed in the (cool enough) photospheric or nonmagnetic  model, which is taken into account in this analysis.
Calculations were made for the model combinations of the spectral types G, K, and M, as plotted in the previous section, and for a range of G, K, and M dwarfs with different underlying models. In Table \ref{tab:overview} the resulting minimum values of the residual intensity, V, and Q are listed for all calculated model combinations and for the following molecular lines: the MgH line at 5197.35\,\AA, the (blended) FeH lines at 9923.44\,\AA\ and 9923.35\,\AA, the CaH line at 6937.03\,\AA, and the (blended) TiO lines at 7056.43\,\AA.

An overall comparison of the obtained signal shows the advantages of different molecules for various spectral type investigations in Fig. \ref{fig:vq}. In general, the Stokes $V$ and $Q$ signals are stronger for cooler spot temperatures because of the stronger molecular lines.
Clear differences can be seen in the usefulness of the analyzed molecules for the different spectral types and for different contrasts of the assumed photosphere and spot temperatures.
In G dwarfs,  the considered  MgH and FeH lines  show the strongest Stokes $V$ and $Q$ signals, while other MgH lines, such as the line at 5201.6\,\AA\ are, in principle,  stronger than FeH, but are blended by atomic lines.  In K dwarfs, MgH shows strong Stokes $V$ signals, but the Stokes $Q$ signals are generally stronger in CaH. For M dwarfs, CaH shows an overall stronger signature and may be more readily detected than TiO (as predicted in \cite{berdyuginaetal2006a} and MgH. TiO itself, however, exhibits strong Stokes $V$ and $Q$ signals especially in M dwarfs. The Stokes $V$ signals for FeH are the strongest in all G dwarfs, with a spot temperature of 3800K and in almost all K dwarfs with this spot temperature. Atomic lines and other molecular lines were not included in the MgH spectral range calculations to solely show Paschen-Back features.
Although large telescopes with high-resolution spectropolarimeters
exist, detecting magnetic fields on cool stars remains a difficult task. To use the available observing time in its full capacity, it is important to efficiently select targets and exposure times upon prior modeling, as presented in this work.


\begin{acknowledgements}
This work is supported by the European Research Council (ERC) Advanced Grant HotMol (ERC-2011-AdG 291659).
\end{acknowledgements}


\bibliographystyle{aa}
\bibliography{journals,nafram}
\end{document}